\documentclass{fairmeta}
\newcommand{\modelname}{MHR}
\author{Aaron Ferguson}
\author{Ahmed~A.~A.~Osman}
\author{Berta Bescos}
\author{Carsten Stoll}
\author{Chris Twigg}
\author{Christoph Lassner}
\author{David Otte}
\author{Eric Vignola}
\author{Fabian Prada}
\author{Federica Bogo}
\author{Igor Santesteban}
\author{Javier Romero}
\author{Jenna Zarate}
\author{Jeongseok Lee}
\author{Jinhyung Park}
\author{Jinlong Yang}
\author{John Doublestein}
\author{Kishore Venkateshan}
\author{Kris Kitani}
\author{Ladislav Kavan}
\author{Marco Dal Farra}
\author{Matthew Hu}
\author{Matthew Cioffi}
\author{Michael Fabris}
\author{Michael Ranieri}
\author{Mohammad Modarres}
\author{Petr Kadlecek}
\author{Rawal Khirodkar}
\author{Rinat Abdrashitov}
\author{Romain Pr\'evost}
\author{Roman Rajbhandari}
\author{Ronald Mallet}
\author{Russell Pearsall}
\author{Sandy Kao}
\author{Sanjeev Kumar}
\author{Scott Parrish}
\author{Shoou-I Yu}
\author{Shunsuke Saito}
\author{Takaaki Shiratori}
\author{Te-Li Wang}
\author{Tony Tung}
\author{Yichen Xu}
\author{Yuan Dong}
\author{Yuhua Chen}
\author{Yuanlu Xu}
\author{Yuting Ye}
\author{Zhongshi Jiang}
\affiliation{Meta}
\date{\today}
\correspondence{\email{mhr@meta.com}}

\metadata[Code]{\url{https://github.com/facebookresearch/MHR}}
\abstract{
We present \modelname, a parametric human body model that combines the decoupled skeleton/shape paradigm of ATLAS with a flexible, modern rig and pose corrective system inspired by the Momentum library. Our model enables expressive, anatomically plausible human animation, supporting non-linear pose correctives, and is designed for robust integration in AR/VR and graphics pipelines.
}

\usepackage{graphicx}
\usepackage{amsmath, amssymb, booktabs, tabularx, comment, wrapfig}
\usepackage[dvipsnames]{xcolor}
\usepackage{subcaption}
\usepackage{balance}

\usepackage{tabularx}
\usepackage{wrapfig}
\usepackage{multirow}
\usepackage{float}
\usepackage{multicol}
\usepackage{cuted}
\usepackage{stfloats}
\usepackage{abstract}
\usepackage{authblk}

\usepackage{pgfplots}
\pgfplotsset{width=8cm,compat=1.9}

\title{\modelname: Momentum Human Rig}

\begin{document}

\twocolumn[{
    \maketitle
    \begin{figure}[H]
    \hsize=\textwidth
    \centering
    \vspace*{-0.2in}
    \includegraphics[trim={6cm 0 6cm 0},clip,width=\textwidth]{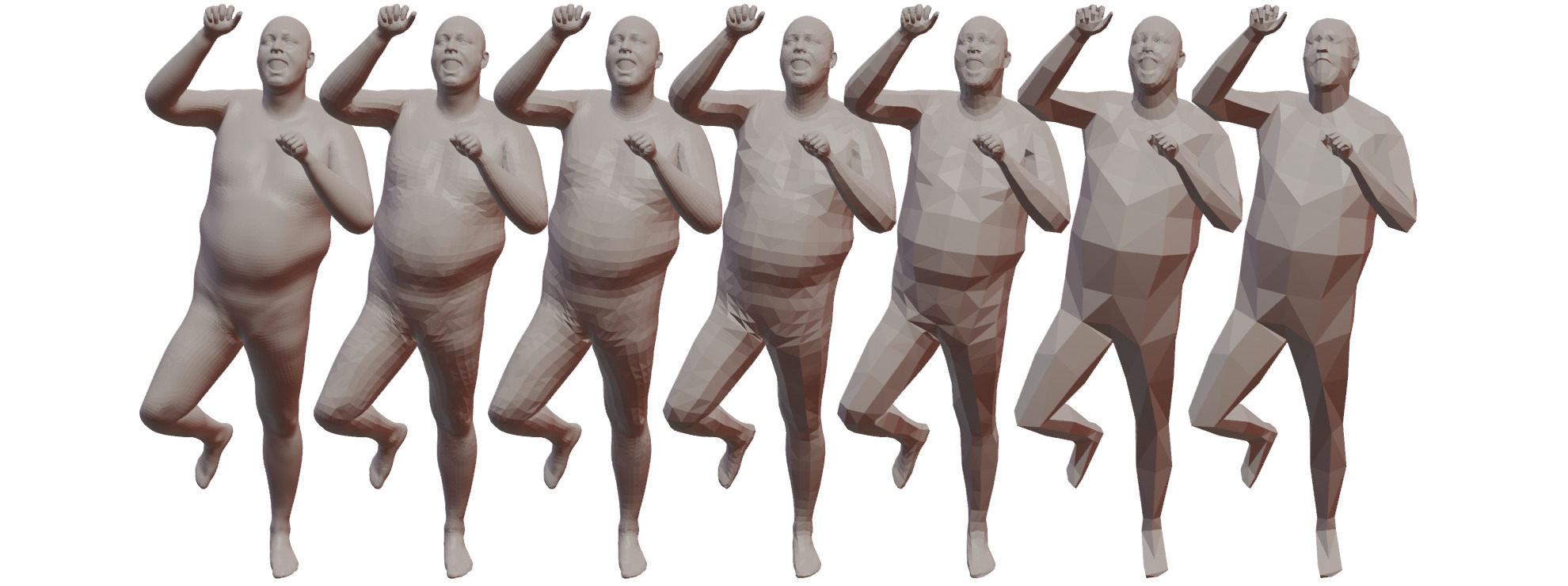}
    \vspace*{-0.1in}
    \caption{\modelname\ provides precise, decoupled control of skeletal and surface attributes at different LODs.}
    \label{fig:proto_lods}
    %\vspace*{-0.1in}
    \end{figure}
}]
%\saythanks

%\maketitle

\section{Introduction}

The landscape of digital human modeling has rapidly evolved, driving progress across fields such as avatar creation~\cite{lombardi2019neural,saito2020pifuhd,weng2022humannerf,xiu2022icon}, motion capture~\cite{yin2023hi4d,cheng2023dna,peng2021neural,zheng2022structured}, simulation of human-object interactions~\cite{bhatnagar2022behave,chao2015hico,wang2024review}, and generative character synthesis~\cite{peng2024charactergen,ren2023make,liang2024rich}. Central to these innovations are parametric body models~\cite{SMPL:2015,SMPL-X:2019,xu2020ghum,anguelov2005scape,wang2020blsm,yang2020facescape}, which translate compact shape and pose descriptors into articulated meshes. The ability to flexibly and accurately represent human form and movement is essential for enabling new applications and deepening our understanding of human-centric data.

The dominant approaches ~\cite{SMPL:2015,SMPL-X:2019,STAR:ECCV:2020,osman2022supr,xu2020ghum} for parametric modeling of the human body focus on personalizing a generic template through linear blendshapes. Their skeletal joints are derived from the surface through weighted sums, and the mesh is driven with linear blend skinning (LBS)~\cite{kavan2007skinning} and pose-dependent corrections. While this paradigm achieves plausible 3D reconstruction, it presents several inherent limitations. Deriving internal skeletal joints from surface vertices introduces incorrect correlations, limiting direct control over skeletal attributes and complicating keypoint fitting. Furthermore, the entanglement of shape and skeleton impedes precise customization of body proportions and soft tissue, which is especially problematic for artists and production pipelines.

ATLAS~\cite{ATLAS:2025} introduces several key innovations to address the limitations of previous parametric human body models. First, it explicitly decouples the external body shape from the internal skeleton, allowing for independent control and customization of both soft-tissue and skeletal attributes. The model is trained at high resolution and features an anatomically motivated skeleton with 77 joints, supporting detailed and realistic articulation. ATLAS further enhances mesh realism by applying sparse, non-linear pose corrective deformations before linear blend skinning, which improves the accuracy of deformations around challenging joints and prevents unwanted correlations between distant body parts. Trained on an extensive dataset of 600,000 high-resolution scans covering a wide range of identities and poses, ATLAS achieves greater expressivity and generalization than prior models. Additionally, it supports robust single-image fitting pipelines by leveraging decoupled shape and skeleton representations, advanced priors, and recent developments in high-fidelity human modeling, resulting in more plausible and accurate reconstructions in diverse scenarios.

However, the potential use of ATLAS in industry is limited by a two factors. First, its expression model based on FLAME~\cite{FLAME:SiggraphAsia2017} is not compatible with most artist-based workstreams which favor sparse, semantic expressions.
%Second, its identity space used CAESAR, whose license can be limiting for certain industrial cases.
And second, its skeleton was not optimized for the addition of pose correctives on top. %Finally, ATLAS was released with a single resolution, which might not be suitable for all computing requirements.

\textbf{Our approach.} To address these challenges, we propose \modelname, an expressive parametric human body model that builds on the ATLAS foundation and introduces key updates for industry and artist needs. \modelname\ features a decoupled skeleton and mesh, semantic expression blendshapes, and a fully compliant identity dataset, all supported at different levels-of-detail. The model is designed for flexibility, expressivity, and legal clarity, supporting a wide range of applications in graphics, vision, and AR/VR. It is released under a clear, industry-friendly license that enables free experimentation.

\section{Related Work}

\noindent\textbf{3D Human Mesh Modeling.}
The modeling of 3D human meshes has evolved significantly over the past two decades. SCAPE~\cite{Anguelov05} pioneered the separation of pose and shape by representing deformations at the triangle level, inspiring a series of works that refined deformation models, improved mesh registration, and extended applications to soft-tissue dynamics~\cite{hasler2009statistical,hirshberg2012coregistration,freifeld2012lie,chen2013tensor,pons2015dyna}. The introduction of SMPL~\cite{SMPL:2015} marked a shift to vertex-based representations, leveraging blendshapes for both shape and pose corrections and employing linear blend skinning (LBS) to articulate the mesh using joints inferred from the surface~\cite{allen2006learning}. Subsequent models such as STAR~\cite{STAR:ECCV:2020}, Frank~\cite{joo2018total}, SMPL-H~\cite{romero2017embodied}, and SMPL-X~\cite{SMPL-X:2019} expanded the modeling space to include hands and faces, introduced more compact or sparse corrective representations, and merged multiple body part models. SUPR~\cite{osman2022supr} and GHUM~\cite{xu2020ghum} further advanced the field by incorporating federated datasets and non-linear shape spaces, respectively. Despite these advances, most of these models regress skeletal joints from the mesh surface, which can entangle shape and skeleton in undesirable ways.

\vspace{1mm}
\noindent\textbf{Skeleton Models.}
In biomechanics, there has been a focus on constructing anatomically faithful skeletons and musculoskeletal systems~\cite{rajagopal2016full,seth2016biomechanical,nitschke2020efficient}, as well as optimizing for internal structures like fat, muscle, and bone~\cite{Dicko2013AnatomyT,gilles2010creating,kadlevcek2016reconstructing,saito2015computational,zhu2015adaptable}. However, these approaches often depend on specialized simulation tools~\cite{lee2009comprehensive} or growth models, making them less accessible for graphics and animation. More recent efforts, such as OSSO~\cite{keller2022osso} and BOSS~\cite{shetty2023boss}, extract anatomical skeletons from SMPL meshes using medical data, but retargeting these skeletons to new poses typically requires additional optimization. Graphics-oriented models that decouple skeleton and shape, like BLSM~\cite{wang2020blsm} and SKEL~\cite{keller2023skin}, offer promising directions but may lack features such as open licensing, expressive pose correctives, or fine-grained finger control. \modelname~and ATLAS~\cite{ATLAS:2025} stands out by enabling direct manipulation of decoupled shape and scale, supporting finger articulation and providing a rich set of pose correctives.

\vspace{1mm}
\noindent\textbf{Pose Corrective Deformations.}
Capturing pose-dependent deformations has been a longstanding challenge. Early solutions applied local vertex offsets near joints to mitigate artifacts like joint collapse~\cite{Lewis:2000:PSD}, while others interpolated between precomputed deformations for key poses~\cite{Allen:TOG:2002,kurihara2004modeling,rhee2006real} or introduced PCA-based corrective spaces for each joint~\cite{kry2002eigenskin}. SMPL~\cite{SMPL:2015} and its derivatives learn mappings from joint rotations to mesh deformations, with STAR~\cite{STAR:ECCV:2020} introducing sparsity and GHUM~\cite{xu2020ghum} leveraging non-linear networks for greater flexibility. However, dense mappings can introduce unwanted correlations across the mesh. \modelname\ and ATLAS aims to balance expressivity and control by employing sparse, non-linear pose correctives that minimize spurious dependencies while accurately modeling complex deformations.

% \vspace{1mm}
% \noindent\textbf{Industry Rigs and Multi-Resolution Parametric Models.}
% \javier{add references and checked, mostly generated by metamate}
% In the animation, VFX, and gaming industries, production rigs are often designed to decouple pose parameterization from the underlying joint structure, allowing for more modular and flexible control. These rigs typically feature explicit skeleton hierarchies that are independent of the mesh topology, enabling artists to retarget animations, swap meshes, or adjust proportions without reauthoring the entire rig. Furthermore, industry-standard rigs frequently support multiple levels of detail (LODs), providing lightweight versions for real-time applications and high-resolution variants for cinematic rendering or data generation. This multi-resolution approach ensures that the same character can be efficiently used across a variety of platforms and production stages. Recent research models, including ATLAS, are increasingly adopting these principles by providing explicit skeleton definitions, decoupled parameter spaces, and scalable mesh resolutions, bridging the gap between academic models and industry requirements.

\section{Model}

In this section, we present the general formulation of the \modelname\ mesh model. This formulation is equivalent to that of ATLAS, but it is included here for completeness.

We formulate our human body model with an explicitly decoupled external surface and an internal skeleton. \modelname\ uses linear blend skinning (LBS)~\cite{kavan2007skinning} following the pose parametrization described in Section~\ref{sec:poseparameterization}. It contains $n_j = 127$ joints, parametrized with $204$ pose parameters (including $68$ skeleton transformation parameters). It supports six different resolutions (levels of detail/LoDs) with $73639$, $18439$, $10661$, $4899$, $2461$, $971$ and $595$ vertices. \modelname\ is formally specified as follows: 
% \vspace{-.5em}
\begin{equation}
X(\beta, \theta) = M(\tilde{X}(\beta^s, \beta^f, \theta), \mathcal{B}^k(\beta^k), \theta, \omega) 
\end{equation}
\vspace{-1em}
\begin{equation}
\tilde{X}(\beta^s, \beta^f, \theta) = \bar{X} + \mathcal{B}^s(\beta^s, \mathcal{S}) + \mathcal{B}^f(\beta^f, \mathcal{F}) + \mathcal{B}^p(\theta, \mathcal{P})
% \vspace{-.25em}
\end{equation}
where the resulting vertices $X \in \mathbb{R}^{3V}$ are a function of input blendshape coefficients $\beta$ and pose parameters $\theta$. The result is obtained by applying the linear blend skinning function $M$ with skinning weights $\omega$ to the neutral template $\bar{X} \in \mathbb{R}^{3V}$. Unlike SMPL, $M$ also uses transformation parameters (joint location offsets or hand isotropic scale) either through an additional set of coefficients $\beta^k$, or raw (i.e. the $68$ parameters previously mentioned).
The template $\bar{X}$ contains identity surface deformations $\mathcal{B}^s(\beta^s, \mathcal{S}) = \sum_{n=1}^{|\beta^s|} \beta^s_n\mathcal{S}_n$, facial expressions $\mathcal{B}^f(\beta^f, \mathcal{F}) = \sum_{n=1}^{|\beta^f|} \beta^f_n\mathcal{F}_n$, and pose correctives $\mathcal{B}^p(\theta, \mathcal{P})$ to account for skinning artifacts caused by LBS~\cite{kavan2007skinning}. Unlike prior works~\cite{SMPL:2015,SMPL-X:2019} that derive skeletal joint centers from this customized identity shape, our mesh at this stage remains unposed, \textit{unscaled}, and aligned to a fixed internal skeleton. This prevents spurious correlations between vertices and joints from affecting the posing.

In the following sections we describe in more detail the pose, expression, identity and pose corrective deformations.

\subsection{Pose Parametrization}
\label{sec:poseparameterization}
\modelname\ is a parametric human body model that combines a  decoupled skeleton and surface bases. 

\textbf{Joints:} The underlying skeleton is built using the Momentum \cite{momentumWEB} library, where each joint is parameterized by 3 translations $T_{t}$ , 3 rotations $T_{rot}$ represented as euler angles in XYZ order, and 1 uniform scale $T_{S}$. Additionally, each joint has a constant translation offset $T_{off}$ relative to its parent joint, and a constant rotation offset $T_{prerot}$ (typically called pre-rotation) that orients the joints local coordinate system. In \modelname\ the pre-rotation is usually set up so that the joints x-axis points in the direction of the bone, and thus rotations around the x-axis are twists. Additionally they are oriented so that rotations around the other axes are symmetric (e.g. a positive rotation around the z-axis will bend the knees backwards).

The complete local-to-world transformation $T_w$ for each joint is calculated as the composition of homogeneus transformations entailed by the components described before:

\begin{align}
T_{w} &= T_{p} \times T_{off} \times T_{t} \times T_{prerot} \times T_{rot} \times T_{s}
\end{align}

\begin{figure}
  \centering
  \includegraphics[trim={3cm 1.5cm 3cm 1cm},clip,width=\linewidth]{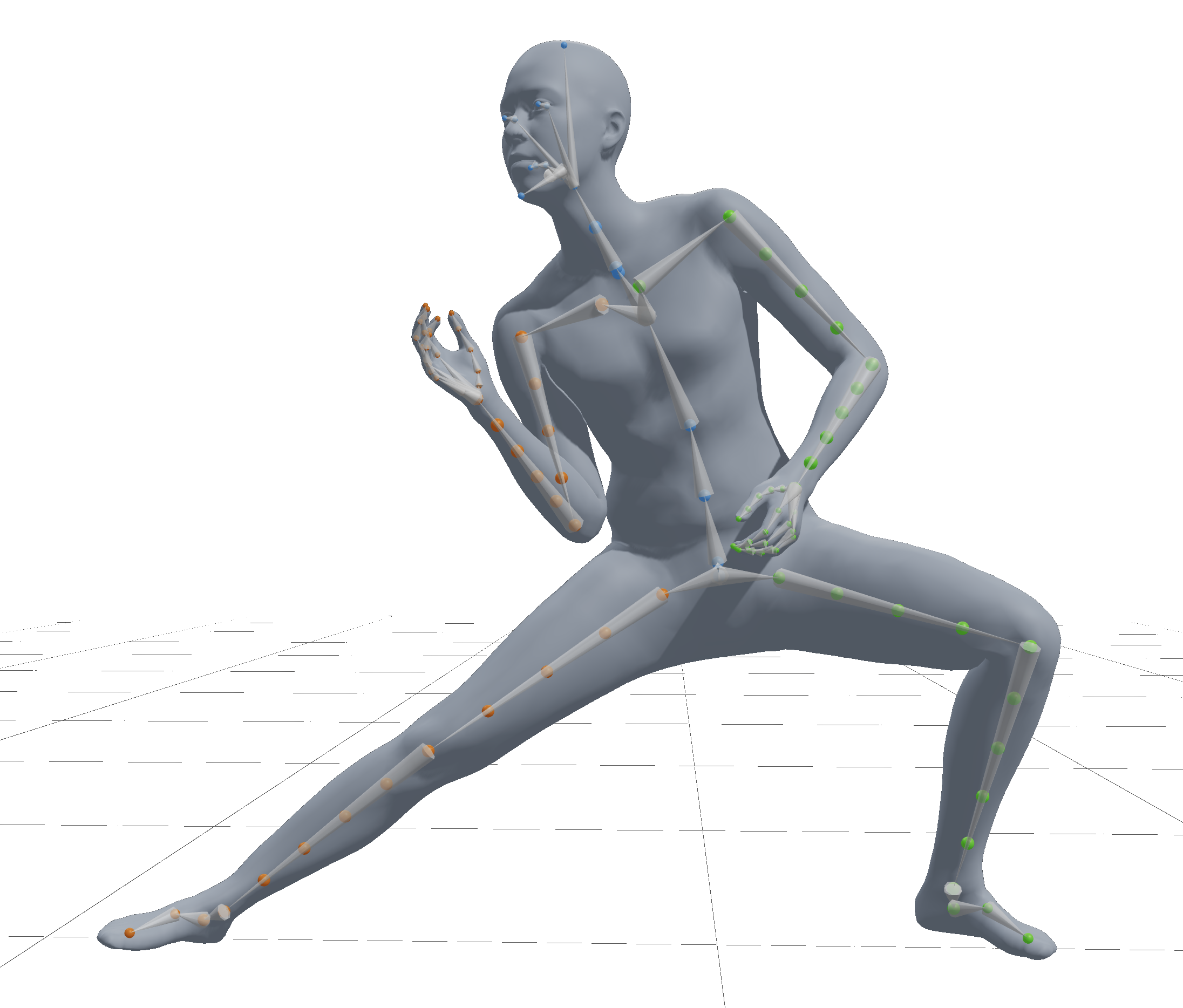}
  \vspace{-0.2in}
  \caption{\textbf{\modelname\ Skeleton}.}
  \label{fig:proto_skeleton}
  \vspace{-0.2in}
\end{figure}

\textbf{Skeleton:} The kinematic hierarchy of the skeleton is stored as a list of $n_j$ joints, with $n_j = 127$ for \modelname. The skeleton can be parameterized with a vector $\Theta_j$ of size $n_j * 7$ that contains the translation, rotation, and scale values for each joint. This pose can be used to articulate the skeleton as seen in Fig.~\ref{fig:proto_skeleton}. This representation allows the full articulation of all degrees of freedom in every joint. In practice we do not want to allow all degrees of freedom in every joint, as most joints will not have translational DoFs, and several, such as the elbow, should only allow rotation around a single axis.

\begin{figure*}
%\vspace{-3mm}
\begin{subfigure}{\textwidth}
\centering
\includegraphics[width=\textwidth]{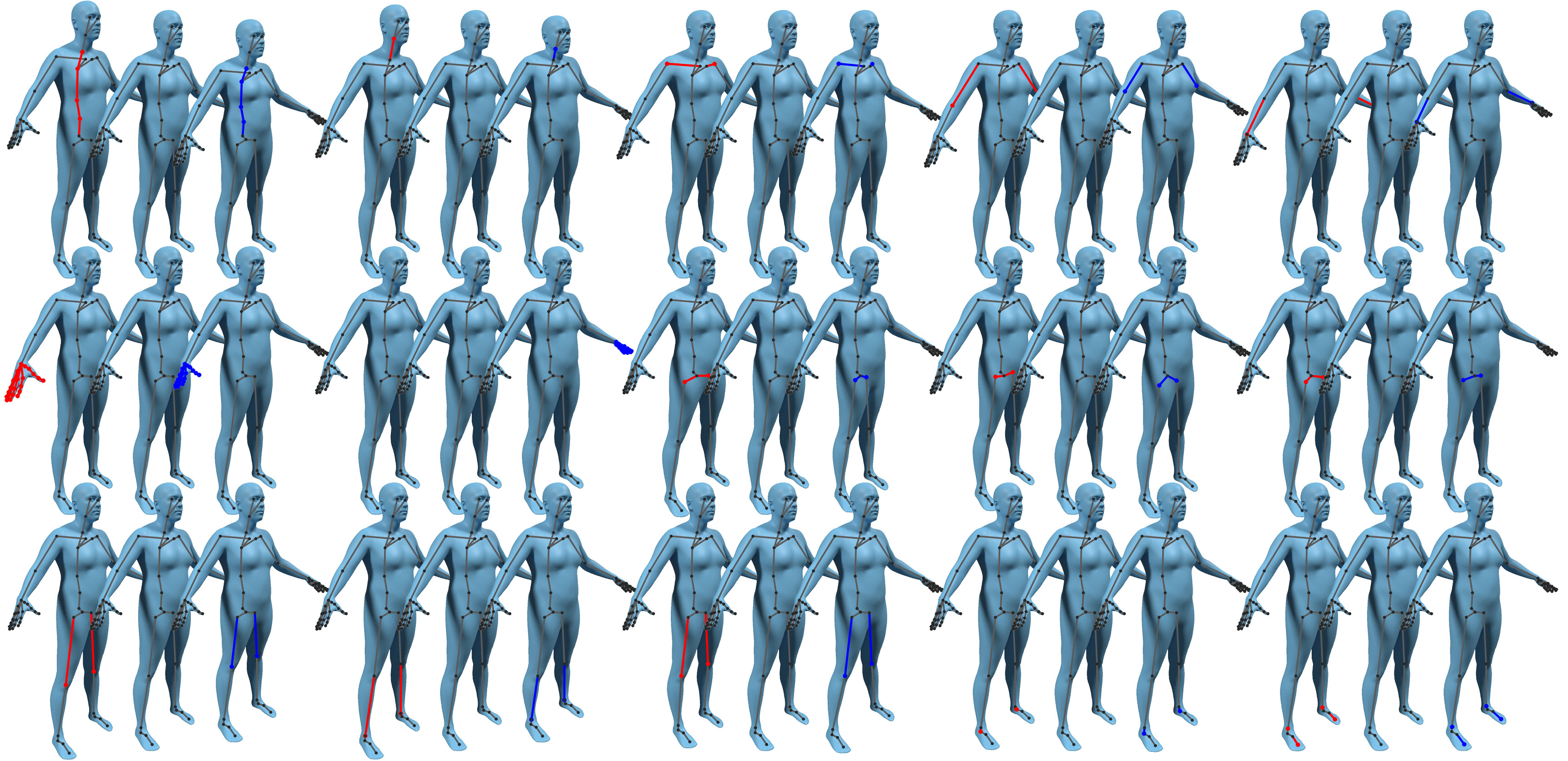}
\caption{\textbf{Full body transformations}. Note that the hand parameters modify the hand isotropic scale instead of its length.}
\end{subfigure}
%\vspace{-3mm}
\begin{subfigure}{\textwidth}
\centering
\includegraphics[width=\textwidth]{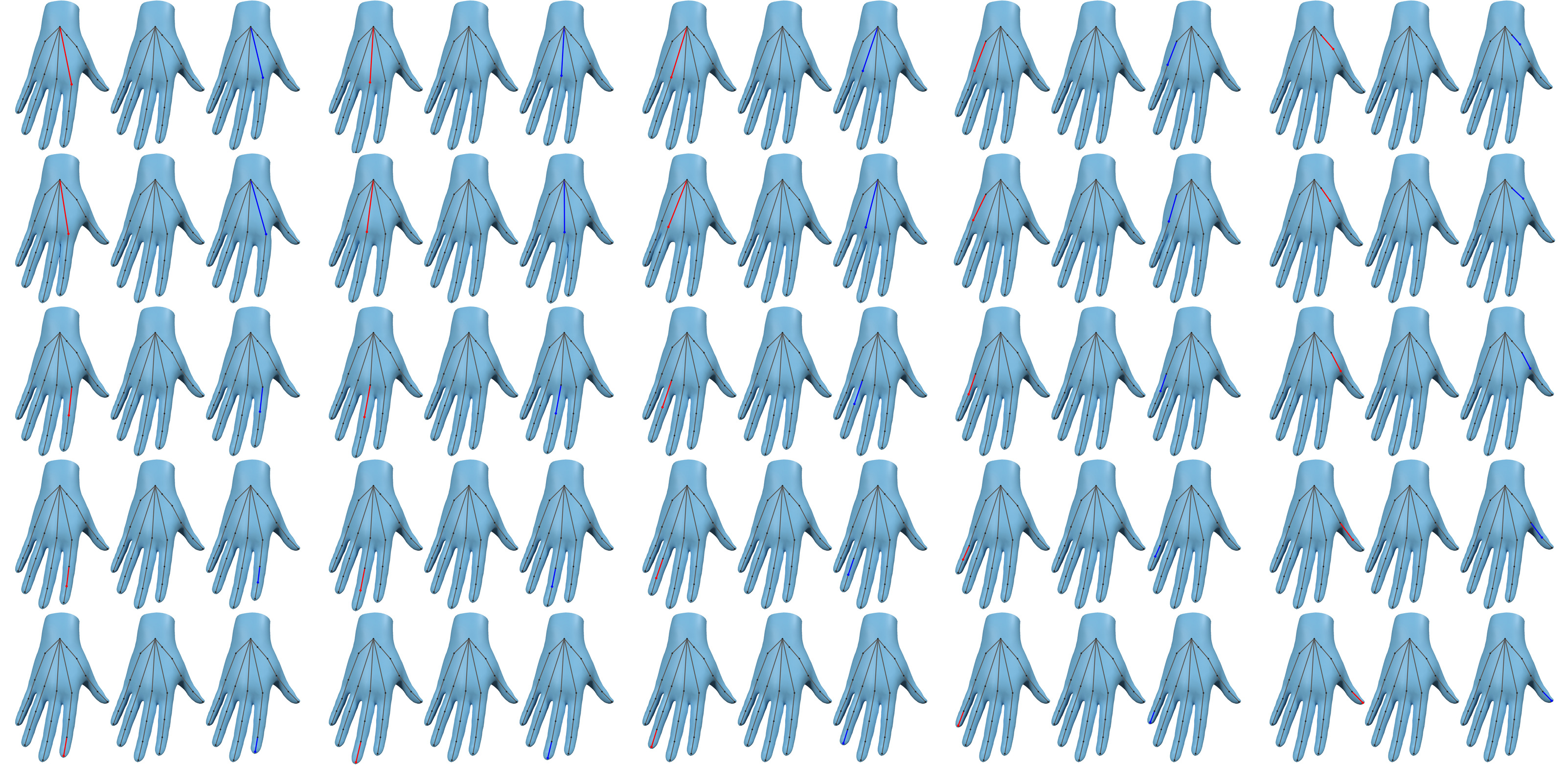}
\caption{\textbf{Hand skeleton transformations}}
\end{subfigure}
%\vspace{-3mm}
\caption{\textbf{\modelname~Skeleton Transformations}. Each cell shows the effect of changing one bone (or set of bones) length. Middle body is neutral, left (right) one shows bone length increase in red (blue).}
%\vspace{-3mm}
\label{fig:skel_transformations}
\end{figure*}

\textbf{Parameters:} To achieve this, we introduce a parameter vector $\Theta_p$, with $n_p = 204$ parameters for \modelname, that is mapped to the joint parameters $\Theta_j$ via a linear transformation $T_p$, i.e. $\Theta_j = T_p * \Theta_p$. This enables several key features that are used heavily in \modelname:

\begin{enumerate}
\item It enables defining a subset of active degrees of freedom by only setting non-zero values to joint parameters we actually want to be able to articulate (e.g. only enable a single rotational DoF of a joint)
\item It allows us to define configurations where a single model parameter influences multiple joint parameters. An example is that \modelname\ reduces the LBS candy wrapper effect by activating supplemental twist joints in the limbs by fractions of the main joint.
%uses a single model parameter for bending the upper spine forward and backward, which is applied in different amounts to multiple spine joints in our model. Other applications are for example rotating supplemental twist joints by fractions of a main joint.
\item It also enables having a single joint parameter being influenced by multiple model parameters. For example \modelname\ parameterizes spine bending with two overlapping parameters for the upper and lower spine.
\end{enumerate}

We split the $204$ parameters into $n_{pose}=136$ pose parameters and $n_{skel}=68$ skeleton transformation parameters (see Fig.~\ref{fig:skel_transformations} for a visualization of the skeleton transformations). The latter are used to define the limb lengths and other skeletal identity parameters and are assumed to be constant for a performer/sequence, while the pose parameters change per frame.

\textbf{Skinning:} \modelname\ uses 4 joint influences per vertex for LoDs 1 to 4
%\javier{should we describe 4 or 6?}
, and 8 joint influences per vertex for LoD 0. Unlike SMPL and ATLAS, we use artist-defined skinning weights without any further optimization. While optimizing skinning weights can reduce training error, we noticed that optimized weights tend to lack structure and locality, crucial components for artist workstreams.

\subsection{Facial Expressions}

Most existing research models (e.g.~\cite{ATLAS:2025, FLAME:SiggraphAsia2017}) use dense, entangled expression spaces derived from data. These expression models present three main advantages. First, they are derived from thousands of scans, which mean they can potentially model nuanced expressions that might not be present in artist-based rigs. Second, their orthogonal spaces make them optimally compact. Third, they effectively model correlations between different parts of the face, which make them easier to optimize. 
On the other hand, they present two critical problems. First, since unposing is an ill-posed, unresolved research problem~\cite{bednarik2024stabilization}, data-driven expression spaces typically contain residual pose variation. Removing this pose variation is critical to model subtle but common gestures like blinking, which should be strictly decorrelated from pose. Second, artist workstreams typically favor semantic, sparse expression spaces.

For this reason, \modelname\ includes expressions that follow the facial coding system (FACs) \cite{Ekman1978FACS} and are sparse and semantic. These $72$ expressions were sculpted by an artist. In our experience, the expression coefficients can be optimized well despite the correlation between them. These expressions simplified substantially the connection of \modelname\ with synthetic data generation pipelines and helped eliminating spurious pose movement in some of our related work.

\begin{figure}
  \centering
  \includegraphics[width=\linewidth]{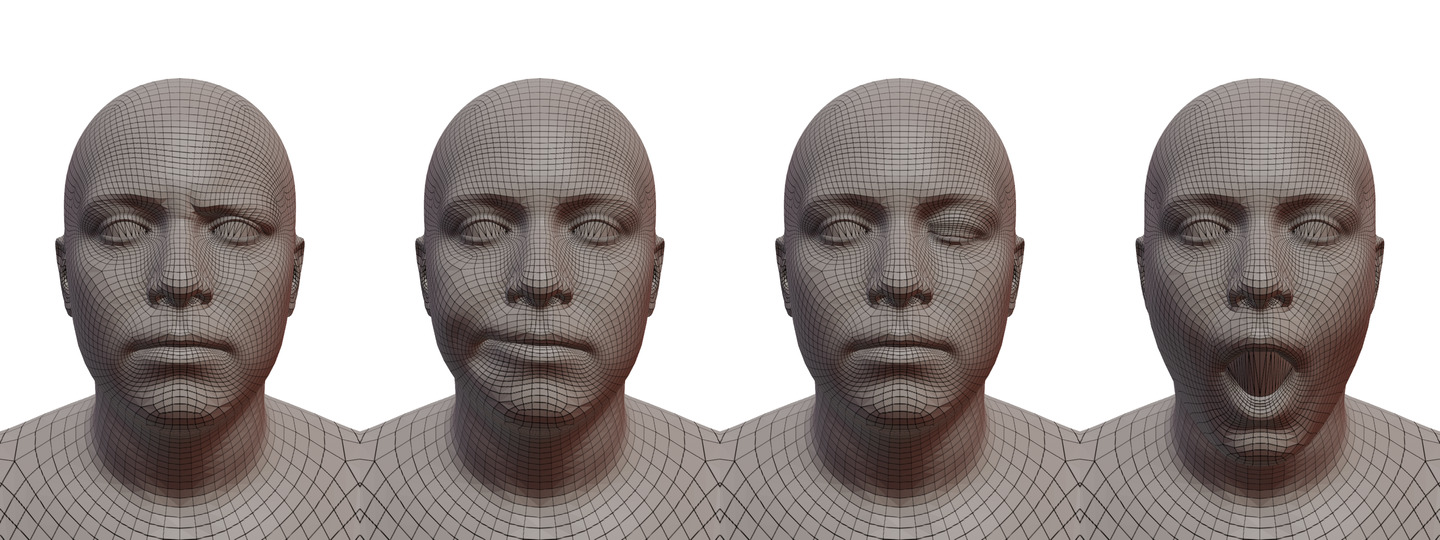}
  \vspace{-0.2in}
  \caption{\textbf{\modelname\ Expression} Example of four \modelname\ expressions fully activated.}
  \label{fig:proto_expressions}
  \vspace{-0.2in}
\end{figure}

\subsection{Identity Space}
The identity space control the intrinsic (i.e. fixed across frames for a particular subject) body shape for a given skeleton structure.
This means that, unlike SMPL, the identity space does not change limb lengths or subjects height.
We define our identity space $\{\beta^s, \mathcal{S}\}$ as the concatenation of three body-specific disjoint identity spaces for the body $\{\beta^{sb}, \mathcal{S}_s\}$, head (or skull) $\{\beta^{ss}, \mathcal{S}_h\}$ and hands $\{\beta^{sh}, \mathcal{S}_h\}$.
This partition of the identity space gives us two main advantages. First, it gives artists better control over shape changes, simplifying the process of achieving a particular look. Second, it allows us to use three different large datasets of part-specifics scans.

For the body, we used %Size NorthAmerica~\cite{sizenaWEB},
a dataset originally composed by $13664$ scans with a wide variety of body shape, age and ethnicity. We filtered the dataset to $7110$ scans by removing unsuitable subjects (e.g. underage or noisy).
%all subjects under the age of 18, subjects captured in Texas or Illinois, and overall noisy scans.
We only used one relaxed pose scan per subject, although adding the rest of the poses could improve the fidelity of the model in underexposed areas like armpits.
We register the data to the \modelname\ topology at LOD1 (i.e. $18439$ vertices), which offers a good balance between detail and compute efficiency.
The registration is performed with non-rigid ICP with a mixture of data and regularization losses. Importantly, we optimize not only our model parameters, but also a set of vertex offsets in neutral pose.
The main data loss is L2 point-to-surface summed over the data vertices.
To make the registration more robust, we also included an L2 keypoint loss that measures the difference between the \modelname\ joints and 3D inferred keypoints.
The 3D keypoints are obtained by rendering the meshes from multiple viewpoints, extracting 2D keypoints, and triangulating them according the the virtual camera 3D calibration.
In terms of regularization, we simply used a joint limit loss that penalizes the square difference between parameters and their limits (defined manually) when the parameters are outside the defined range. 
We registered the dataset multiple times, creating after each iteration an identity space by running PCA on the neutral template (offsets plus current identity blendshapes). In the first iteration, the identity space was initialized with three artist-defined blendshapes depicting a high- and low-BMI female and male characters.

Given that the hands and head in the full body dataset are not very high quality, we used separate datasets to model those body parts. For the hands, we used an internal dataset of hand-specific scans obtained form $3000$ subjects. Those scans were registered with a non-linear ICP pipeline similar to the body pipeline. 
To model the head identity, we extended the collection of head captures in~\cite{martinez2024codec} to a total of $2138$ subjects. 
We modified the Pixel Codec Avatar~\cite{ma2021pixel} to take identity conditioning and fitted the model to our dataset. For the purpose of \modelname\ we are interested in the neutral (unposed, no expression) geometry of each subject fitted in this process.

\begin{figure}
    %\hsize=\textwidth
    \centering
    \includegraphics[trim={16cm 1cm 19cm 3cm},clip,width=\columnwidth]{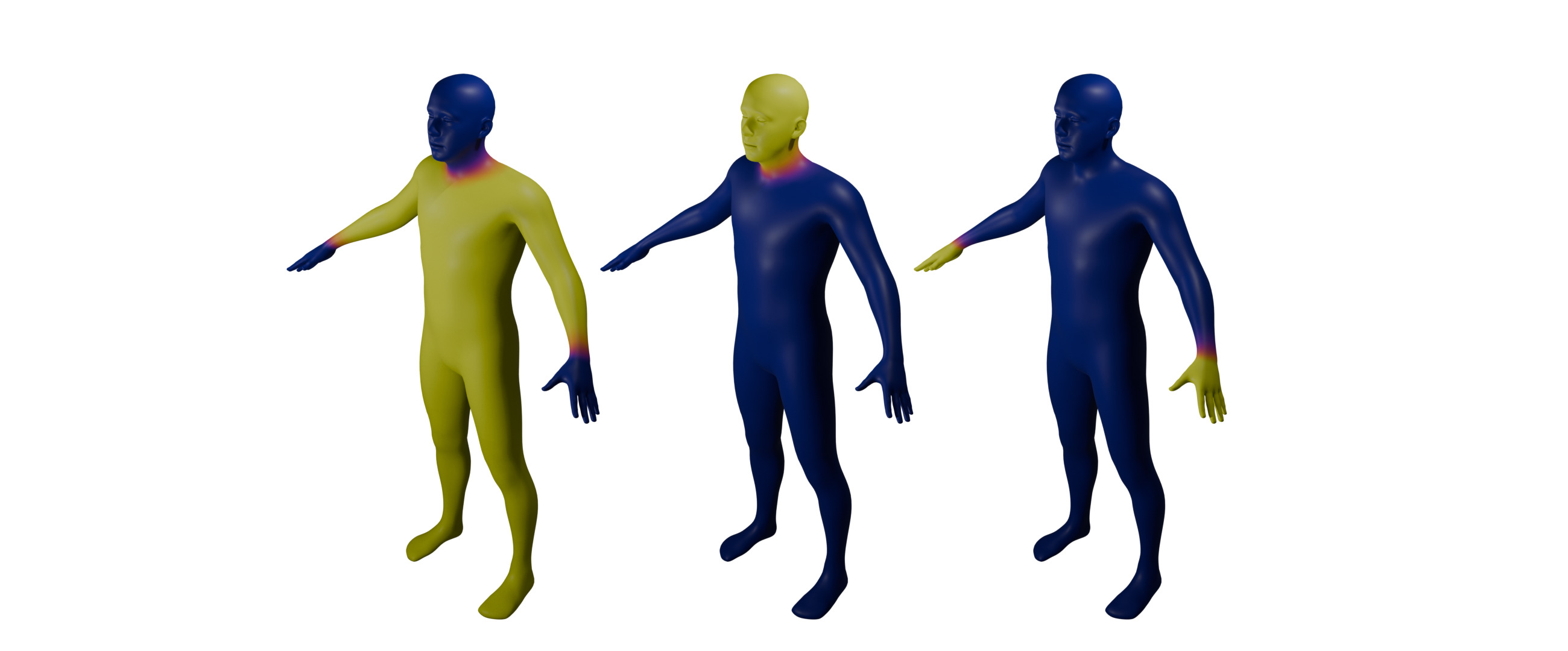}
    \caption{Body, head and hand masks for partitioned identity shape space}
    \label{fig:partitioned_pca_space}
\end{figure}

In order to obtain a smooth identity space without breaks between the body parts, we multiplied the data in each of the subsets by a soft mask depicted in Figure~\ref{fig:partitioned_pca_space} and ran PCA on each of the weighted subsets separately. 
To increase the amount of data available in each of the datasets, we mirrored the available scans before training our models. As a positive side-effect, the spurious asymmetries contained in the registrations are compressed into two specific components in body and head subspaces, which were removed from our model.
We selected an empirical number of components per body part ($20$ body, $20$ head and $5$ hand components) and concatenated them to obtain our final \modelname\ identity space. The mean shape in \modelname\ is the sum of the means in each (weighted) subspace. 

\begin{figure*}[t]
  \centering
  \includegraphics[width=\linewidth]{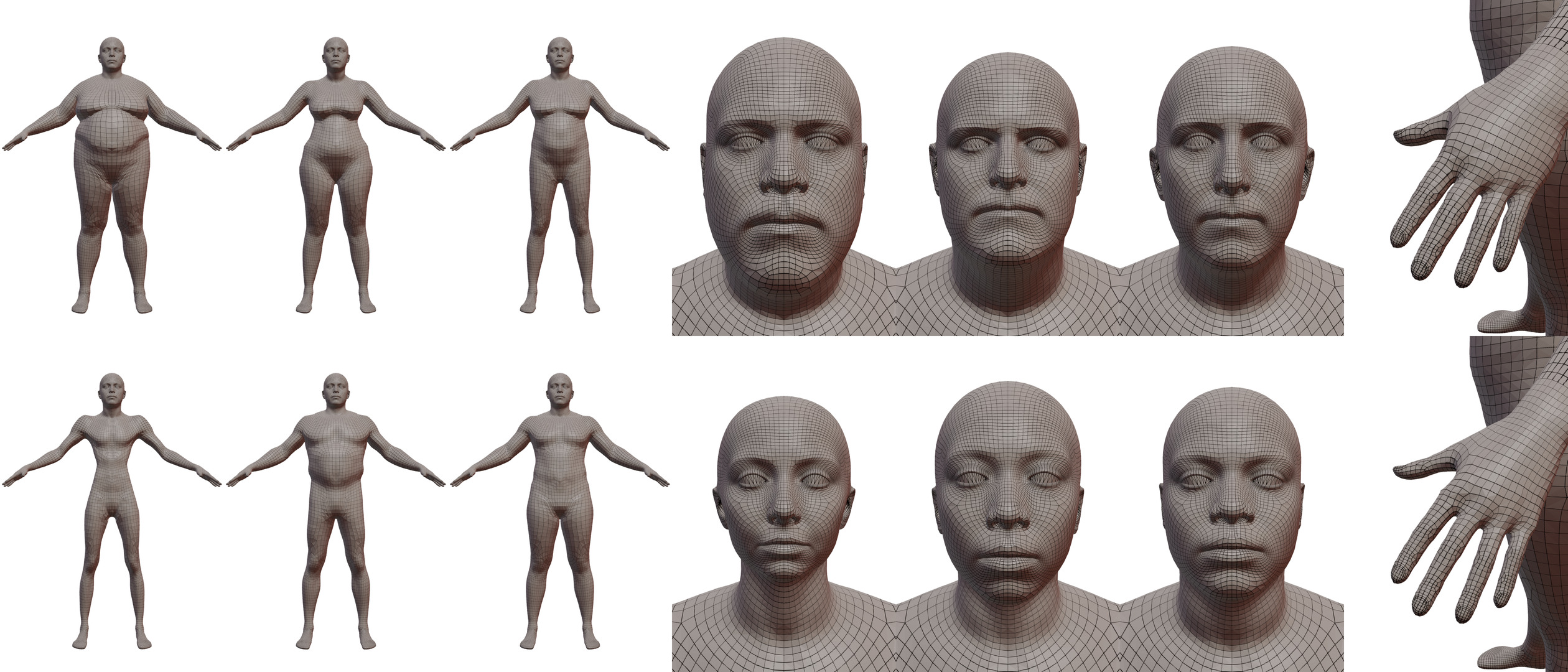}
  \vspace{-0.2in}
  \caption{\textbf{\modelname\ Identity Space} First three body components, three face components, and one hand component ($3$ and $-3$ standard deviations from the mean at the top and bottom row respectively).}
  \label{fig:proto_expressions}
  \vspace{-0.2in}
\end{figure*}

\subsection{Pose Correctives}

\begin{figure}[h]
  \centering
  \includegraphics[width=\linewidth]{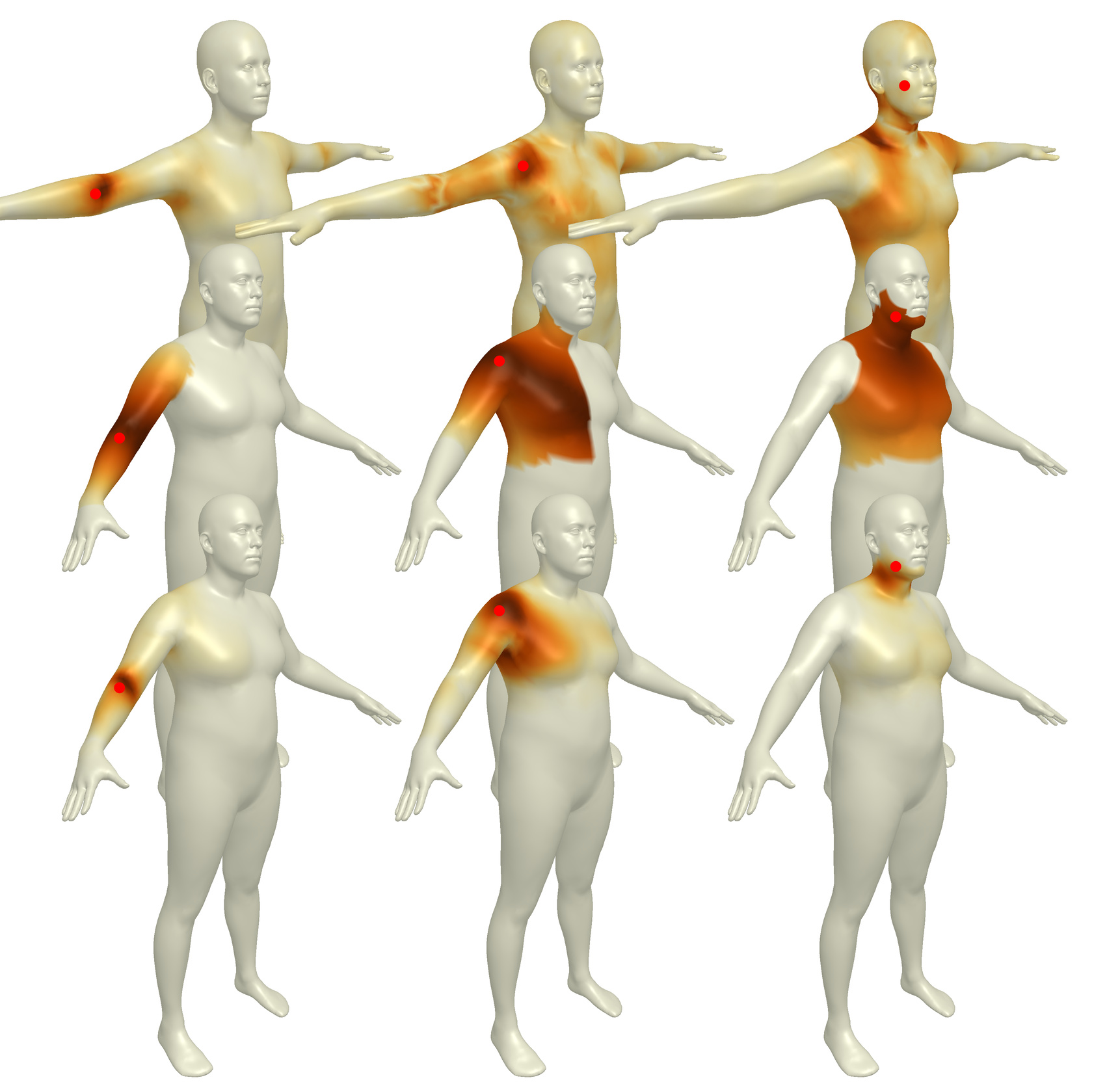}
  \caption{\textbf{Sparse Pose Correctives.} The first row displays pose correctives from SMPL-X. The second row shows the inverse geodesic initialization for our pose corrective activations, and the third row demonstrates their sparsity after convergence.}
  \label{fig:pcb_trained}
\end{figure}

We train pose correctives on $26000$ scans ($13000$ full body, $13000$ hand scans) at LOD1 following the ATLAS definition of pose correctives. We add the formulation here for completeness.
For the LBS posed mesh to look realistic, pose-dependent deformations prior to LBS~\cite{kavan2007skinning} are critical. Our correctives function $\mathcal{B}^p(\theta, \mathcal{P}) \in \mathbb{R}^{6J} \rightarrow{} \mathbb{R}^{3V}$ takes joint angles in 6D~\cite{zhou2019continuity} and outputs vertex offsets. While prior work demonstrates the strength of sparse-linear~\cite{STAR:ECCV:2020,osman2022supr} and dense-non-linear ~\cite{xu2020ghum} pose correctives, we converge these directions. As non-linear operations inevitably couple the inputs and complicate sparsity enforcement, we decompose $\mathcal{B}^p$ into a local, non-linear operation and a sparse, geodesic-initialized linear operation. First, we write the local, non-linear operation as:
\begin{equation}
    \text{Non-Linear}_j(\theta) = \text{MLP}\left( \{R_{6d}(\theta_a) - R_{6d}(\vec{0}) \mid a \in n(j)\} \right)
\end{equation}
The local operation $\text{Non-Linear}_j(\theta)$ processes joint $j$ and its immediate neighbors $n(j)$. Here, $R_
{6d}(\theta_a) - R_{6d}(\vec{0})$ represents the 6D rotational deviation from the identity rotation for joint $a$. A lightweight MLP processes each joint $j$ together with its adjacent parent and child joints, producing a $c$-dimensional embedding that encodes their poses. As we will regularize the extent of vertices this joint group centered at $j$ will affect, this local joint group entanglement effectively enables non-linear expressivity while avoiding spurious joint-vertex correlations. Finally, the pose corrective for a joint $j$ is:
\begin{equation}
    \mathcal{B}^p_j = \phi(A_j) \odot \left(P_j \times \text{Non-Linear}_j(\theta)\right)
\end{equation}
Following STAR~\cite{STAR:ECCV:2020}, $\phi$ represents the ReLU activation applied to joint mask $A_j \in \mathbb{R}^V$, $P_j \in \mathbb{R}^{3V \times c}$ is the pose corrective weight, and $\times$ is standard matrix multiplication. $\left(P_j \times \text{Non-Linear}_j(\theta)\right)$ yields the non-linear pose dependent mesh deformations, with $\phi(A_j)$ enforcing vertex deformation sparsity per joint. For vertex $i$, we initialize the $i$-th element of $A_j$ as $(1 - d(i, j)) \boldsymbol{1}_{i \in \text{seg}(j)}$, where $d(i, j)$ is the normalized geodesic distance from vertex $i$ to the vertex ring around $j$, and $\boldsymbol{1}_{i \in \text{seg}(j)}$ indicates if vertex $i$ belongs to joint $j$'s corresponding or adjacent body part. This initialization, coupled with L1 regularization on $\phi(A)$, encourages sparsity in activation. Figure \ref{fig:pcb_trained} shows the activation mask pre- and post-training, showing pose correctives concentrated around the actuated joint.

\subsection{Implementation Details}

The provided model is implemented using the Momentum library~\cite{momentumWEB}, which provides an efficient C++/Python APIs for rig definition, parameter transforms, and skinning. Models can be loaded and exported in artist-friendly formats like Autodesk FBX and GLTF. The model can be easily integrated in pytorch neural network frameworks. 

As previously mentioned, pose correctives and identity models are trained at LOD1. We transfer the obtained blendshapes (identity, expression, and last layer of corrective MLPs) to the rest of the LODs. For lower LODs, we perform a linear mapping based on closest face and barycentric coordinates, while for LOD0 we subdivide the LOD1 correctives to achieve a smoother result.

\section{Evaluation}

To assess our approach, we utilize the 3DBodyTex~\cite{3DBodyTex} dataset, which comprises high-resolution scans of $100$ male and $100$ female subjects, each captured in two distinct poses. To measure the expressiveness of each model, we optimize both body shape and pose parameters by minimizing the sum of the distances between each point on the scans to the closest point on the model surface. Since hands, face and hair are not reliable in the scans, we manually masked out their corresponding vertices and added a keypoint term which measures the distance between the provided head and hand landmarks and the corresponding model joints. We minimize the sum of those two losses with Adam for 2500 iterations with a learning rate of 0.01.

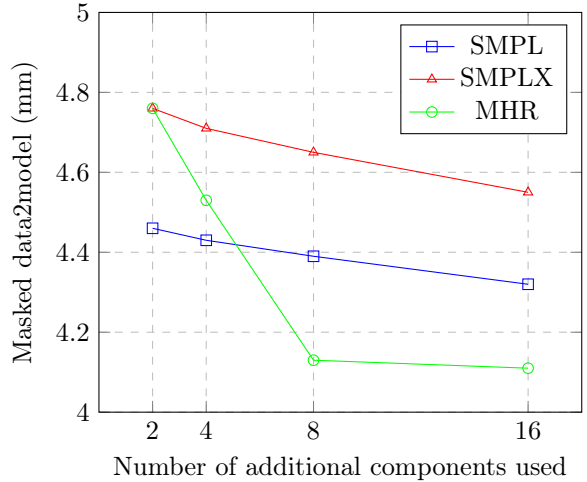
\begin{figure}
\begin{tikzpicture}%[width=\linewidth]
\begin{axis}[
    ylabel={Masked data2model (mm)},
    xlabel={Number of additional components used},
    xmin=0, xmax=18,
    ymin=4, ymax=5,
    xtick={2,4,8,16},
    legend pos=north east,
    ymajorgrids=true,
    xmajorgrids=true,
    grid style=dashed,
]

\addplot[
    color=blue,
    mark=square,
    ]
    coordinates {
    (2.0,4.46)(4.0,4.43)(8.0,4.39)(16.0,4.32)
    %(2.0,5.94)(4.0,5.52)(8.0,5.10)(16.0,4.74)(35.0,4.46)(37.0,4.43)(41.0,4.39)(49.0,4.32)
    };
    \legend{SMPL}
    \addplot[
    color=red,
    mark=triangle,
    ]
    coordinates {
    (2.,4.76)(4.,4.71)(8.,4.65)(16.,4.55)
    %(2.,6.28)(4.,5.73)(8.,5.33)(16.,5.12)(35.,4.76)(37.,4.71)(41.,4.65)(49.,4.55)
    };
    \addlegendentry{SMPLX}
    \addplot[
    color=green,
    mark=o,
    ]
    coordinates {
    (2.,4.76)(4.,4.53)(8.,4.13)(16.,4.11)
    };
    \addlegendentry{\modelname}
\end{axis}
\end{tikzpicture}
  \caption{\textbf{Quantitative Evaluation on 3DBodyTex.} We report vertex-to-vertex error (mm) with different numbers of fitting components. Note use SMPL and SMPL-x include $33$ additional components to account for the additional pose components in \modelname}
  \label{fig:3dbodytex_quant}
\end{figure}

\begin{figure*}[h]
  \centering
  \includegraphics[width=\linewidth]{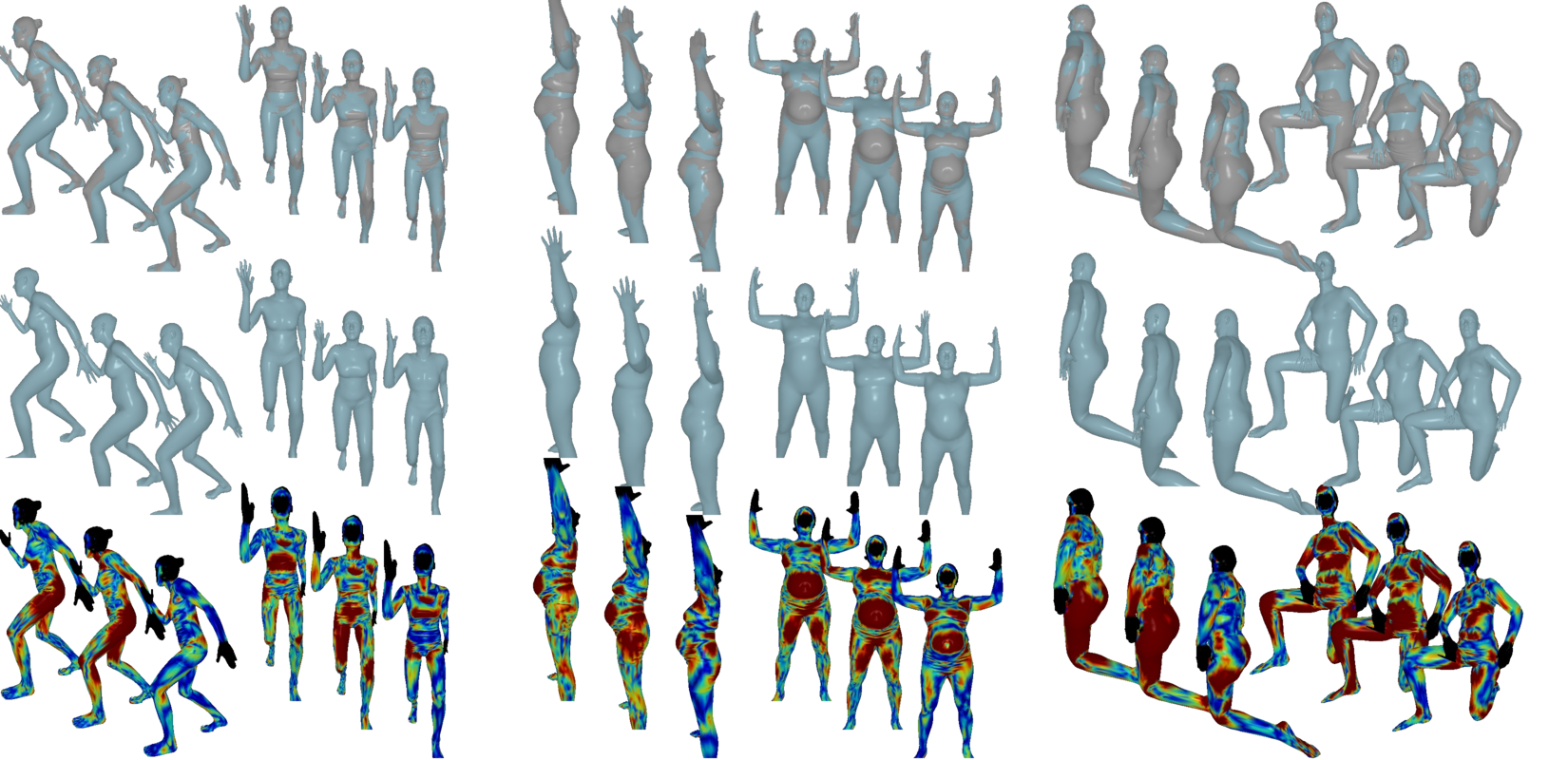}
  \vspace{-0.2in}
  \caption{\textbf{Qualitative Results on 3DBodyTex.} We visualize two views of three different scans in 3DBodyTex. In each column, we see from left to right SMPL, SMPL-X and \modelname. First row shows the overlap of the scan and model estimation, the second one shows only the model, and the third one the distance from model to scan as a heatmap (masked out areas in black).}
  \label{fig:3dbodytex}
  \vspace{-0.2in}
\end{figure*}

We report the average distance from scan points to the closest model surface excluding face, hair and hands in Figure~\ref{fig:3dbodytex_quant}, while qualitative outcomes are illustrated in Figure~\ref{fig:3dbodytex}. Our model demonstrates a lower fitting error with fewer components, confirming \modelname's capability to represent posed human body shapes for previously unseen identities. Qualitatively, our model excels particularly at the extremities of articulated joints (such as elbows and knees) and provides a closer fit to the target scan’s shoulders.

% \begin{figure}[h]
%   \centering
%   \vspace{-0.2in}
%   \includegraphics[width=\linewidth]{figures/eval_proto.png}
%   \vspace{-0.2in}
%   \caption{\textbf{Quantitative Evaluation on 3DBodyTex.} We report vertex-to-vertex error (mm) with different numbers of fitting components. For ATLAS, we report the combination of the number of shape and scale components used.}
%   \label{fig:3dbodytex_quant}
%   \vspace{-0.2in}
% \end{figure}

\section{Discussion and Future Work}

The creation of a parametric model requires a number of decision that trade off different characteristics like accuracy, universality, or ease of artistic use, among others. 

The skeleton definition is one of such decisions. The decoupling between joints and parameters described in Section~\ref{sec:poseparameterization} gives us the flexibility of defining complex skeletons driven by compact parameterizations. In \modelname\ we decided to simplify the ATLAS skeleton, removing some additional joints in the glutes and upperback. Those additional joints increase the accuracy of LBS-only models, but can make pose optimization harder. Given that we are releasing a model with pose-correctives, we decided to err on the side of simplicity for the skeleton. 

One critical component in real-time rigs is the maximum amount of joints that influence any given vertex. While LBS accuracy and smoothness can benefit from a larger limit, we decided to favor a strict limit of four joint influences per vertex in all but the highest L0 LOD, since there the four joint limit resulted in sharp creases around the skinning boundaries.

\modelname\ focuses on fitting and modeling geometry as observed in minimal scans. Unlike FLAME or SMPL-X, \modelname\ does not include explicit eyeball geometry; we plan to include this in future iterations of the model. We also plan to add an explicit mouth system similar to~\cite{rasras2024lips} that models teeth and tongue. 

The facial expression and pose corrective blendshapes are independent from body shape in~\modelname. Pose correctives become more accurate when they vary across body shapes~\cite{STAR:ECCV:2020}, and shape-dependent expression models can also improve realism~\cite{vlasic2006face}. We will explore in future work how to condition pose correctives and expression on body shape.

There is a number of other future directions that we plan to explore with \modelname, including integration of soft-tissue and clothing models, real-time optimization and deployment in AR/VR pipelines, as well as extending it to stylized characters.

\section{Conclusions}

\modelname\ advances the state of the art in parametric human modeling by combining the decoupled skeleton/shape paradigm of ATLAS with a modern, corrective-driven rig. The result is a flexible, expressive model suitable for animation, vision, and AR/VR applications.

%\newpage

\balance

\bibliographystyle{assets/plainnat}
\bibliography{main.bib}

\clearpage
\newpage
\beginappendix
\section{Author Contributions}
\label{appendix:contributors}

\modelname\ is the culmination of over 9 years of work on multiple iterations of anatomically inspired body models developed at Meta. The following people contributed to the model (or earlier iterations):
\begin{description}
\item \textbf{Aaron Ferguson} - Modeling, rigging, and artistic supervision
\item \textbf{Ahmed~A.~A.~Osman} - Statistical model development, scan registrations, evaluation
\item \textbf{Berta Bescos} - Head scan registrations
\item \textbf{Carsten Stoll} - Model parameterization, Momentum library development and support
\item \textbf{Chris Twigg} - Momentum library development and support
\item \textbf{Christoph Lassner} - Software and pipeline development
\item \textbf{David Otte} - Tech art supervision
\item \textbf{Eric Vignola} - Rigging, tooling, and artistic supervision.
\item \textbf{Fabian Prada} - ATLAS development
\item \textbf{Federica Bogo} - Statistical model development, scan registrations, pipeline development
\item \textbf{Igor Santesteban} - Testing, QA
\item \textbf{Javier Romero} - Testing, statistical model development, scan registrations
\item \textbf{Jenna Zarate} - Technical program management
\item \textbf{Jeongseok Lee} - Momentum library development and support
\item \textbf{Jinhyung Park} - Statistical body model, pose correctives, SAM3D integration, ATLAS development
\item \textbf{Jinlong Yang} - SMPL conversion tooling
\item \textbf{John Doublestein} - Infrastructure and pipeline lead, modeling, rigging and testing
\item \textbf{Kishore Venkateshan} - Software and pipeline development, rig and model evaluation and improvement
\item \textbf{Kris Kitani} - Open sourcing, SAM3D integration, ATLAS development
\item \textbf{Ladislav Kavan} - Body pose correctives
\item \textbf{Marco Dal Farra} - Technical program management
\item \textbf{Matthew Hu} - Hand scan registrations
\item \textbf{Matthew Cioffi} - Expression blendshapes, modeling and rigging
\item \textbf{Michael Fabris} - Rigging and modeling
\item \textbf{Michael Ranieri} - Software development support
\item \textbf{Mohammad Modarres} - Head shape modeling, archetype modeling
\item \textbf{Petr Kadlecek} - Software and pipeline development
\item \textbf{Rawal Khirodkar} - ATLAS development
\item \textbf{Rinat Abdrashitov} - Software and pipeline development, hand model improvements
\item \textbf{Romain Pr\'evost} - Head scan registration
\item \textbf{Roman Rajbhandari} - Technical Artist
\item \textbf{Ronald Mallet} - Technical program management
\item \textbf{Russell Pearsall} - Previous rig development
\item \textbf{Sandy Kao} - Tech art supervision
\item \textbf{Sanjeev Kumar} - Testing,  QA
\item \textbf{Scott Parrish} - Rigging and modeling
\item \textbf{Shouu-I Yu} - ATLAS development
\item \textbf{Shunsuke Saito} - ATLAS development
\item \textbf{Takaaki Shiratori} - ATLAS development
\item \textbf{Te-Li Wang} - Head model development
\item \textbf{Tony Tung} - Technical program management
\item \textbf{Yichen Xu} - ATLAS development
\item \textbf{Yuan Dong} - Model parameterization, evaluation and QA
\item \textbf{Yuhua Chen} - Testing, QA
\item \textbf{Yuanlu Xu} - Momentum library development and support
\item \textbf{Yuting Ye} - Model parameterization, QA, Momentum library support, open sourcing
\item \textbf{Zhongshi Jiang} - Model and UV improvements
\end{description}

% \clearpage
% \newpage
% \input{model_figures}

\end{document}